\documentclass[12 pt]{article}
\usepackage{graphicx}
\begin{document}
\baselineskip=24pt
\title{\bf{DGLAP equation and $x$-evolution of Non-singlet structure function in Leading order}}
\author{D K Choudhury \thanks{E-mail: dkc\_physics@yahoo.co.in} \\Department of Physics\\Gauhati University\\Guwahati-781014,India \and  Pijush Kanti Dhar \thanks{E-mail: dr\_pijush@yahoo.co.in}\\HOD, Physics\\Arya Vidyapeeth College\\Guwahati-781016, India}
\maketitle
\begin{abstract}
Study of parton distribution function (PDF) is a topic of significant interest in QCD. To overcome the shortcomings of conventional PDFs, several alternative methods have been suggested in recent years. The present work reports the $x$-distribution of the non-singlet structure function using the complete solution of Taylor-approximated DGLAP equation. The formalism is more general than the similar ones reported in recent literature. Our predictions are compared with the CCFR neutrino data as well as exact results.



\end{abstract}

\newpage

\section{Introduction:}
Recently \cite{DKCPKD} we have reported the $t$-evolution of the non-singlet structure functions at low and high $x$ as the complete solution of Taylor-approximated DGLAP equation \cite{GRIBOVLIPA,LIPATOV,DOKSH,AP,GALTA}. The present paper reports the prediction of corresponding $x$-evolution. Although the conventional wisdom on DGLAP evolution does not favour such a feature, such a possibility was reported \cite{JKSDKCGKM} sometime back. The main reason is that in this approach, the DGLAP equation is converted into a first order differential equation in two variables $x$ and $t$ ( $t=\log \frac{Q^2}{\Lambda^2}$) instead of only $t$ and has $x$ evolution as its natural  outcome.

There are, of course, various methods that have been developed recently like the  brute-force \cite{MIYAMAKUMANO,HIKUMI}, Laguerre-polynomial \cite{KOBKONKUM,RTO} and Mellin-transformation \cite{BLU1,BLU2,MSTWVO,KUNAGA,NAGAKU} etc. for numerical solution of DGLAP equation. All these methods have good numerical accuracy and computational efficiency, particularly the Mellin-transformation one.  All these solutions are very useful in the studies of a top-down scenario in order to determine the origin of ultra-high energy cosmic rays. In addition, the matrix technique to get the numerical solution of the DGLAP evolution equation has also been proposed \cite{PGRAT}, which provides very good precision.

In recent years, the study of parton distribution function (PDF) has itself become a topic of significant interest. The standard approach to the problem is based on a choice of a specific functional form in $x$ with a few parameters and to obtain fit of them by comparing the exact solution of the evolution equation with global data. Based on this approach, several sets of pdf have been produced in recent years \cite{SIALEKHIN,JPUMPLIN,MRST11,MRST22,MRSTDURHAM}. More recently, however, shortcoming of such an approach has been pointed out : the form of parametrization is itself a source of bias. To overcome this limitation, numerical method incorporating the notion of neural network \cite{SFORTEETAL,SFORTE,LDDETAL,LDPEHHIO,ROJOLA} has been suggested and pursued. Alternative conventional models like statistical \cite{RNB11,RNB22,JSOFFER} and chiral quark model \cite{HDMMG} have also received cosiderable attention in this respect. 
            
The present paper reports the analytical $x$-evolution of the non-singlet structure function in the formulation of Ref. \cite{DKCPKD} which is much more general than that of earlier works \cite{JKSDKCGKM} and compares with the other approaches. Since such an analytical solution is in good agreement with experimental data, hence this type of study bears important significance. 
 In Sec \S \ref{secchap4:formalism}, we develop the essential formalism while in Sec \S \ref{secchap4:Results}, we will discuss the results. Sec \S \ref{secchap4:Comments and conclusions} will summarize a conclusion.

\section{FORMALISM:}
\label{secchap4:formalism}
\subsection{Particular Solution of DGLAP equation at low $x$:}
\label{subsecchap4:Soln at low x}
The DGLAP evolution equation for non-singlet structure function in standard form is given by \cite{AP} 
\begin{equation}
\label{eqn:ch4eqA}
\frac{\partial F^{NS}\left(x,Q^2\right)}{\partial \log Q^2}=P_{NS}\left(x,Q^2\right)\otimes F^{NS}\left(x,Q^2\right)
\end{equation}
 where $x$ is the usual Bjorken variable %/already defined in eq. (\ref{eqn:ch1eq1}) i.e.
\begin{equation}
\label{eqn:ch4eqB}
{x=\frac{Q^2}{2P.q}}\, 
\end{equation}
and $Q^2$ is the four momentum transfer in deep inelastic scattering \cite{HALZMARTIN} process.
The symbol $\otimes$ stands for the usual Mellin convolution in the first variable defined as
\begin{equation}
\label{eqn:ch4eqC}
 {a(x)\otimes f(x)=\int\limits_x^1\,\frac{dy}{y}a(y)f\left(\frac{x}{y}\right)}.
\end{equation}
The non-singlet kernel $P_{NS}(x,Q^2)$ has perturbative expansion in orders of the strong coupling constant $\alpha_s (Q^2)$ as
\begin{equation}
\label{eqn:ch4eqD}
 {P_{NS}(x,Q^2)=\frac{\alpha_s(Q^2)}{2\pi}P_{NS}^{(0)}(x)+\left(\frac{\alpha_s(Q^2)}{2\pi}\right)^2 P_{NS}^{(1)}(x)+.....}
\end{equation}
where $P_{NS}^{(0)}(x)$ and $P_{NS}^{(1)}(x)$ are the one loop \cite{AP} and two loop \cite{FURPETRONZIO,FURPETRONZIO1,FURPETRONZIO2,CURCIFURPE} contributions to the non-singlet structure function respectively.
In case of LO, we neglect two loop contribution and so eq. (\ref{eqn:ch4eqA}) can be rewritten as
\begin{equation}
\label{eqn:ch4eqE}
\frac{\partial F^{NS}\left(x,Q^2\right)}{\partial \log Q^2}=\frac{\alpha_s(Q^2)}{2\pi}P_{NS}^{(0)}(x)\otimes F^{NS}\left(x,Q^2\right)
\end{equation}
Since the evolution of the non-singlet structure function is independent of the gluon, hence one can write from earlier equation,
\begin{equation}
\label{eqn:ch4eqF}
\frac{\partial F^{NS}\left(x,Q^2\right)}{\partial \log Q^2}=\frac{\alpha_s(Q^2)}{2\pi}P_{qq}^{(0)}(x)\otimes F^{NS}\left(x,Q^2\right)
\end{equation}
where we have \cite{KUNAGA,RKELLIS} %/as defined in eq. (\ref{eqn:ch1eq1})
\begin{equation}
\label{eqn:ch4eqG}
{P_{qq}^{(0)}(x)=C_F\left[\frac{2}{(1-x)_+}-1-x+\frac{3}{2}\delta(1-x)\right]}  
\end{equation}
 where $C_F$ is given by $C_F=\frac{(N_{C}-1)}{2N_{C}}$ and $\frac{1}{(1-x)_+}$ is defined by 
\begin{equation}
\label{eqn:ch4eqG1}
\displaystyle {\int\limits_{0}^{1} dx \frac{f(x)}{(1-x)_+}}=\displaystyle {\int\limits_{0}^{1} \frac{f(x)-f(1)}{(1-x)}}.
\end{equation}
 Here $f(x)$ is an arbitrary function and $N_C$ is the number of colours.                           

Using this explicit form of the splitting function $P_{qq}^{(0)}(x)$, eq. (\ref{eqn:ch4eqF}) can be written as \cite{LFAbbot}
\begin{eqnarray}
\label{eqn:ch4eq1A}
 \frac{\partial F^{NS}(x,t)}{\partial t}&=&\frac{\alpha_{s}(t)}{2\pi}\left[\frac{2}{3}\{3+4\log(1-x)\}F^{NS}(x,t)\right.\nonumber\\
& &\left. -\frac{4}{3}\int\limits_x^1\frac{dz}{1-z}\left\{(1+z^2)F^{NS}\left(\frac{x}{z},t\right)
-2F^{NS}(x,t)\right\}\right]
\end{eqnarray}
Here $t=\log\left(\frac{Q^2}{\Lambda ^2}\right)$ and the running coupling constant in Leading order (LO) is  
$\alpha_s(t)=\frac{4\pi}{\beta_0 t}$

with

$\beta_0=\frac{33-2N_f}{3}$, $N_f$ being the number of quark flavours.

This expression can be re-written as
\begin{eqnarray}
\label{eqn:ch4eq1}
 \frac{\partial F^{NS}(x,t)}{\partial t}&=&\frac{A_f}{t}\left[\{3+4\log(1-x)\}F^{NS}(x,t)\right.\nonumber\\
& &\left. +2\int\limits_x^1\frac{dz}{1-z}\left\{(1+z^2)F^{NS}\left(\frac{x}{z},t\right)
-2F^{NS}(x,t)\right\}\right]
\end{eqnarray}
Here  $A_f=\frac{4}{3\beta_0}=\frac{4}{33-2N_f}$ and $\frac{\alpha_{s}(t)}{3\pi}=\frac{A_f}{t}$.

For small $x$ $(x\ll 1)$, it is justified to neglect higher derivatives of $\frac{\partial F^{NS}(x,t)}{\partial x}$ so that $F^{NS}\left(\frac{x}{z},t\right)$ occured in the R.H.S. of eq. (\ref{eqn:ch4eq1}) can be written as
\begin{equation}
\label{eqn:ch4eq2}
{F^{NS}\left(\frac{x}{z},t\right)=F^{NS}(x,t)+ x\sum_{k=1}^\infty u^k\frac{\partial F^{NS}(x,t)}{\partial x}}
\end{equation}
where $u=1-z.$ 

Putting eq. (\ref{eqn:ch4eq2}) in eq. (\ref{eqn:ch4eq1}) and performing the u-integration yields the Taylor-approximated DGLAP equation as
\begin{equation}
\label{eqn:ch4eq3}
{Q(x,t)\frac{\partial F^{NS}(x,t)}{\partial t}+P(x,t)\frac{\partial F^{NS}(x,t)}{\partial x}= R(x,t)F^{NS}(x,t)}
\end{equation}
where $P(x,t)$, $Q(x,t)$ and $R(x,t)$ are explicitly mentioned in of Ref. \cite{DKCPKD} as
\begin{equation}
\label{eqn:ch4eq3A}
{P(x,t)=\frac{-A_f x\left[2\log\left(\frac{1}{x}\right)+(1-x^2)\right]}{t}=\frac{P_1(x)}{t}}
\end{equation}
\begin{equation}
\label{eqn:ch4eq3B}
{Q(x,t)=1}
\end{equation}
and
\begin{equation}
\label{eqn:ch4eq3C}
{R(x,t)=\frac{A_f[3+4\log(1-x)+(x-1)(x+3)]}{t}=\frac{R_1(x)}{t}}
\end{equation}

Eq. (\ref{eqn:ch4eq3}) is frequently referred to as Lagrange's equation \cite{Sneddon,FAYRES}. Its general solution is obtained by solving the following auxiliary systems of ordinary differential equations
\begin{equation}
\label{eqn:ch4eq3D}
\frac{dx}{P(x,t)}=\frac{dt}{Q(x,t)}=\frac{dF^{NS}(x,t)}{R(x,t)F^{NS}(x,t)}
\end{equation}
which subsequently gives
\begin{equation}
\label{eqn:ch4eq3E}
\frac{dx}{P_{1}(x)}=\frac{dt}{Q_{1}(t)}=\frac{dF^{NS}(x,t)}{R_{1}(x)F^{NS}(x,t)}
\end{equation}
where $P_{1}(x)=tP(x,t)$, $Q_{1}(t)=tQ(t)$ and $R_{1}(x)=tR(x,t)$.
   The most general form of the solution of the Taylor-approximated DGLAP equation (eq. (\ref{eqn:ch4eq3})) is \cite{DKCPKD}, 
\begin{equation}
\label{eqn:ch4eq4}
{v=\alpha u^{n(x,t)}+{\beta}}
\end{equation}
where $n(x,t)$ is any real function of $x$ and $t$ and $\beta$,\,$\alpha$ are two arbitrary constants. $u$ and $v$ are two explicit functions of $x$ and $t$ given by \cite{DKCPKD},
\begin{equation}
\label{eqn:ch4eq4A}
{u(x,t)=tX^{NS}(x)}
\end{equation}
and
\begin{equation}
\label{eqn:ch4eq4B}
{v(x,t)=F^{NS}(x,t)Y^{NS}(x)}
\end{equation}

which are two independant solutions of eq. (\ref{eqn:ch4eq3E}).

In order to develop our formalism, we take the numerical value of $n(x,t)=n_{num}=n$ and use the following two physically plausible boundary conditions \cite{DKCATRI}
 \begin{equation}
\label{eqn:ch4eq5}
{F^{NS}(x,t)=F^{NS}(x_0,t)}
\end{equation}
for some low $t=t_0$ and 
\begin{equation}
\label{eqn:ch4eq6}
{F^{NS}(1,t)=0}
\end{equation}
Putting the boundary conditions (\ref{eqn:ch4eq5}) and (\ref{eqn:ch4eq6}) in eq. (\ref{eqn:ch4eq4}), along with expressions for $u(x,t)$ and $v(x,t)$ from eq. (\ref{eqn:ch4eq4A}) and eq. (\ref{eqn:ch4eq4B}) respectively, one gets
\begin{equation}
\label{eqn:ch4eq7}
{F^{NS}(x,t)Y^{NS}(x)=\alpha t^{n}\{[X^{NS}(x)]^{n}-[X^{NS}(1)]^{n}\}}
\end{equation}
Defining
\begin{equation}
\label{eqn:ch4eq8}
{F^{NS}(x_0,t)Y^{NS}(x_0)=\alpha t^{n}\{[X^{NS}(x_0)]^{n}-[X^{NS}(1)]^{n}\}}
\end{equation}
where $X^{NS}(x)$ and $Y^{NS}(x)$ are given as \cite{DKCPKD}

\begin{equation}
\label{eqn:ch4eq8A}
{X^{NS}(x)=\exp\left[-\int\frac{dx}{P_1(x)}\right]}
\end{equation}
and
\begin{equation}
\label{eqn:ch4eq8B}
{Y^{NS}(x)=\exp\left[-\int\frac{R_1(x)}{P_1(x)}dx\right]}
\end{equation} 

This results in
\begin{equation}
\label{eqn:ch4eq9}
{F^{NS}(x,t)=F^{NS}(x_0,t)\frac{Y^{NS}(x_0)}{Y^{NS}(x)}\frac{[X^{NS}(x)]^{n}-[X^{NS}(1)]^{n}}{[X^{NS}(x_0)]^{n}-[X^{NS}(1)]^{n}}}
\end{equation}
As we have \cite{DKCPKD}
\begin{equation}
\label{eqn:ch4eq10}
{X^{NS}(1)\approx 0},
\end{equation}
hence eq. (\ref{eqn:ch4eq9}) yields for $n>0$,
\begin{equation}
\label{eqn:ch4eq11}
{F^{NS}(x,t)=F^{NS}(x_0,t)\frac{Y^{NS}(x_0)}{Y^{NS}(x)}\frac{[X^{NS}(x)]^{n}}{[X^{NS}(x_0)]^{n}}}
\end{equation}
On the other hand for $n<0$, the R.H.S of eq. (\ref{eqn:ch4eq9}) will involve inverse of $X^{NS}(1)$ which is singular and ill-defined, and hence excluded on physical grounds.
Putting the values of $X^{NS}(x)$ and $Y^{NS}(x)$, we get $x$ distribution obtained from complete solution at low $x$,
\begin{equation}
\label{eqn:ch4eq12}
{F^{NS}(x,t)=F^{NS}(x_0,t)\exp\left[\,\,\displaystyle{\int\limits_{x_0}^{x}\left\{\frac{-n}{P_1(x)}+\frac{R_1(x)}{P_1(x)}\right\}dx}\right]}.\;\;\;\;   (n>0)
\end{equation}
where $P_{1}(x)$ and $R_{1}(x)$ are defined as \cite{DKCPKD}
\begin{equation}
\label{eqn:ch4eq13}
{P_{1}(x)=-A_{f} x\left[2\log\left(\frac{1}{x}\right)+(1-x^{2})\right]}
\end{equation}
\begin{equation}
\label{eqn:ch4eq14}
{R_{1}(x)=A_{f}[3+4\log(1-x)+(x-1)(x+3)]}
\end{equation}

Eq. (\ref{eqn:ch4eq12}) is also true for quark of each flavour individually i.e.
\begin{equation}
\label{eqn:ch4eq15}
{\lim_{x\to 0}q_i(x,t)=q_i(x_0,t)\exp\left[\,\,\displaystyle{\int\limits_{x_0}^{x}\left\{\frac{-n}{P_1(x)}+\frac{R_1(x)}{P_1(x)}\right\}dx}\right]},\,\,\,   (n>0)
\end{equation}
Combining $t$-evolution for quark \cite{DKCPKD} at low $x$, viz, 

\begin{equation}
\label{eqn:ch4eq15A}
{q(x,t)=q(x,t_0)\left(\frac{t}{t_0}\right)^{n} \;\;\;\;(n>0)},
\end{equation}

we obtain
\begin{equation}
\label{eqn:ch4eq16}
{\lim_{x\to 0}q_i(x,t)=q_i(x_0,t_0)\left(\frac{t}{t_0}\right)^{n}\exp\left[\,\,\displaystyle{\int\limits_{x_0}^{x}\left\{\frac{-n}{P_1(x)}+\frac{R_1(x)}{P_1(x)}\right\}dx}\right]},\,\,\,   (n>0)
\end{equation}
containing $t$-evolution as well.
Eq. (\ref{eqn:ch4eq12}) and eq. (\ref{eqn:ch4eq16}) are more general than the similar $x$ distribution reported recently \cite{RAJSARMAEUR}.

\subsection{Particular Solution of DGLAP equation at high $x$:}
\label{subsecchap4:Soln. at high x}
For high $x$, the most general form of the solution of the Taylor-approximated DGLAP equation is \cite{DKCPKD}
\begin{equation}
\label{eqn:ch4eq17}
{v^{\prime}=\alpha^{\prime} u^{{\prime}^{p(x,t)}}+{\beta}^{\prime}}
\end{equation}
where $p(x,t)$ is another function of $x$ and $t$ similar to $n(x,t)$ in eq. (\ref{eqn:ch4eq4}).
As in case of low $x$, we assume  $p(x,t)=p_{num}=p$ so as to yield
\begin{equation}
\label{eqn:ch4eq18}
{F^{NS}(x,t)=F^{NS}(x_0,t)\frac{Y^{{\prime}^{NS}}(x_0)}{Y^{{\prime}^{NS}}(x)}\frac{[X^{{\prime}^{NS}}(x)]^{p}-[X^{{\prime}{NS}}(1)]^{p}}{[X^{{\prime}^{NS}}(x_0)]^{p}-[X^{{\prime}{NS}}(1)]^{p}}}
\end{equation}
where $X^{{\prime}^{NS}}(x)$ and $Y^{{\prime}^{NS}}(x)$ are defined as \cite{DKCPKD}

\begin{equation}
\label{eqn:ch4eq18A}
{X^{{\prime}^{NS}}(x)=\exp\left[-\int\frac{dx}{P_1^{\prime}(x)}\right]}
\end{equation}
and
\begin{equation}
\label{eqn:ch4eq18B}
{Y^{{\prime}^{NS}}(x)=\exp\left[-\int\frac{R_1(x)}{P_1^{\prime}(x)}dx\right]}
\end{equation}

As $X^{{\prime}^{NS}}(1)$ is singular \cite{DKCPKD}, for $p>0$, R.H.S of eq. (\ref{eqn:ch4eq18}) is ill-defined and excluded on physical grounds.
If on the other hand, for $p<0$, R.H.S of eq. (\ref{eqn:ch4eq18}) involves $\left[X^{{\prime}^{NS}}(1)\right]^{-1}$, which vanishes identically. Under this condition ($p<0$),
\begin{equation}
\label{eqn:ch4eq19}
{F^{NS}(x,t)=F^{NS}(x_0,t)\frac{Y^{{\prime}^{NS}}(x_0)}{Y^{{\prime}^{NS}}(x)}\frac{[X^{{\prime}^{NS}}(x)]^{p}}{[X^{{\prime}^{NS}}(x_0)]^{p}}}
\end{equation}
Putting the values of $X^{{\prime}^{NS}}(x)$ and $Y^{{\prime}^{NS}}(x)$, we finally get the $x$-distribution obtained from complete solution at high $x$ to be
\begin{equation}
\label{eqn:ch4eq20}
{F^{NS}(x,t)=F^{NS}(x_0,t)\exp\left[\,\,\displaystyle{\int\limits_{x_0}^{x}\left\{\frac{-p}{P_1^{\prime}(x)}+\frac{R_1(x)}{P_1^{\prime}(x)}\right\}dx}\right]}.\;\;\;\;   (p<0)
\end{equation}
where
\begin{equation}
\label{eqn:ch4eq21}
{ P_{1}^{\prime}(x)=\frac{2}{3}A_{f} x(x-1)(x^2+x+4)}
\end{equation}
Eq. (\ref{eqn:ch4eq20}) is also true for each quark flavour individually.
\begin{equation}
\label{eqn:ch4eq22}
{\lim_{(1-x) \to 0}q_i(x,t)=q_i(x_0,t)\exp\left[\,\,\displaystyle{\int\limits_{x_0}^{x}\left\{\frac{-p}{P_1^{\prime}(x)}+\frac{R_1(x)}{P_1^{\prime}(x)}\right\}dx}\right]}.\;\;\;\;   (p<0)
\end{equation}
 Combining $t$-evolution for quark \cite{DKCPKD} at high $x$, viz, 

\begin{equation}
\label{eqn:ch4eq22A}
{q(x,t)=q(x,t_0)\left(\frac{t}{t_0}\right)^{p} \;\;\;\;(p<0)},
\end{equation}
 
we obtain
\begin{equation}
\label{eqn:ch4eq23}
{\lim_{(1-x)\to 0}q_i(x,t)=q_i(x_0,t_0)\left(\frac{t}{t_0}\right)^{p}\exp\left[\,\,\displaystyle{\int\limits_{x_0}^{x}\left\{\frac{-p}{P_1^{\prime}(x)}+\frac{R_1(x)}{P_1^{\prime}(x)}\right\}dx}\right]}.\;\;\;\;  (p<0)
\end{equation}
containing $t$-evolution as well.

\subsection{General Solution of DGLAP equation at low and high $x$:}
\label{subsecchap4: General soln. at low and high x}

In  Subsection  \ref{subsecchap4:Soln at low x} and  Subsection \ref{subsecchap4:Soln. at high x} , we considered the numerical values of $n(x,t)$ and $p(x,t)$ that differentiated the low $x$ from the high $x$ regimes in case of $x$-evolution. Now we attempt to find out the  general and complete analytical solution of DGLAP equation using the explicit functional forms of $n(x,t)$ and $p(x,t)$ and their $x\rightarrow 1$ and $x\rightarrow x_0$ limits at low and high $x$ respectively.

Now in case of low $x$, considering the functional form of $n(x,t)$ and proceeding as in Subsection  \ref{subsecchap4:Soln at low x}, we get
\begin{equation}
\label{eqn:ch4eq24A}
{F^{NS}(x,t)Y^{NS}(x)=\alpha\{[tX^{NS}(x)]^{n(x,t)}-[tX^{NS}(1)]^{n(1,t)}\}}     
\end{equation}

Defining
\begin{equation}
\label{eqn:ch4eq24B}
{F^{NS}(x_0,t)Y^{NS}(x_0)=\alpha\{[tX^{NS}(x_0)]^{n(x_0,t)}-[tX^{NS}(1)]^{n(1,t)}\}}     
\end{equation}
we get,

\begin{equation}
\label{eqn:ch4eq24C}
{F^{NS}(x,t)=F^{NS}(x_0,t)\frac{Y^{NS}(x_0)}{Y^{NS}(x)}\frac{[tX^{NS}(x)]^{n(x,t)}-[tX^{NS}(1)]^{n(1,t)}}{[tX^{NS}(x_0)]^{n(x_0,t)}-[tX^{NS}(1)]^{n(1,t)}}}.   
\end{equation}

Now proceeding as in Subsection \ref{subsecchap4:Soln at low x}, we ultimately get the  general $x$-distribution at low $x$,

\begin{equation}
\label{eqn:ch4eq24}
{F^{NS}(x,t)=F^{NS}(x_0,t)\frac{Y^{NS}(x_0)}{Y^{NS}(x)}\frac{t^{n(x,t)}}{t^{n(x_0,t)}}\frac{[X^{NS}(x)]^{n(x,t)}}{[X^{NS}(x_0)]^{n(x_0,t)}}}.\;\;\;\; (n(1,t)>0)     
\end{equation}
instead of eq. (\ref{eqn:ch4eq11}).

Similarly, considering the functional form of $p(x,t)$ and proceeding as in Subsection \ref{subsecchap4:Soln. at high x}, we ultimately get,

\begin{equation}
\label{eqn:ch4eq25}
{F^{NS}(x,t)=F^{NS}(x_0,t)\frac{Y^{{\prime}^{NS}}(x_0)}{Y^{{\prime}^{NS}}(x)}\frac{t^{p(x,t)}}{t^{p(x_0,t)}}\frac{[X^{{\prime}^{NS}}(x)]^{p(x,t)}}{[X^{{\prime}^{NS}}(x_0)]^{p(x_0,t)}}}.\;\;\;\; (p(1,t)<0)
\end{equation}
instead of  eq. (\ref{eqn:ch4eq19}).

Eq. (\ref{eqn:ch4eq24}) and eq. (\ref{eqn:ch4eq25}) can be re-expressed as
\begin{eqnarray}
\label{eqn:ch4eq26}
F^{NS}(x,t) = F^{NS}(x_0,t)\exp\left[\,\,\displaystyle{\int\limits_{x_0}^{x}\frac{R_1(x)}{P_1(x)}dx}\right]\frac{t^{n(x,t)}}{t^{n(x_0,t)}}  & &  \nonumber  \\
 \times \frac{\exp\left[-n(x,t)\displaystyle{\int\frac{1}{P_1(x)}dx}\right]}{\exp\left[-n(x,t)\displaystyle{\int\frac{1}{P_1(x)}dx}\right]}_{x={x_0}}.\nonumber\\  
 \nonumber\\  (n(1,t)>0)
\end{eqnarray}

and

\begin{eqnarray}
\label{eqn:ch4eq27}
F^{NS}(x,t) =  F^{NS}(x_0,t)\exp\left[\,\,\displaystyle{\int\limits_{x_0}^{x}\frac{R_1(x)}{P_1^{\prime}(x)}dx}\right]\frac{t^{p(x,t)}}{t^{p(x_0,t)}} & & \nonumber  \\
 \times \frac{\exp\left[-p(x,t)\displaystyle{\int\frac{1}{P_1^{\prime}(x)}dx}\right]}{{\exp\left[-p(x,t)\displaystyle{\int\frac{1}{P_1^{\prime}(x)}dx}\right]}_{x=x_0}}.\nonumber\\  
 \nonumber\\  (p(1,t)<0)
\end{eqnarray}

 We now put forward the  generalized statement as: {\bf The most general solution of the Taylor-approximated DGLAP equation at low and high $x$ are obtainable in terms of two arbitrary real functions $n(x,t)$ and $p(x,t)$ such that $n(1,t)$ is greater than zero and $p(1,t)$ is less than zero for any $t$. In case, $n(x,t)$ and $p(x,t)$ have only mild variation in $t$, then $n(1,t) \approx n(1)> 0$ and $p(1,t) \approx p(1)< 0$ can be considered as the $x \rightarrow 0$ and $(1-x)\rightarrow 0$ limit of an interpolating function $H(x,t)$ to be determined empirically from  data.}

\subsection{Alternate Solution of DGLAP equation at low and high $x$:}
\label{subsecchap4:Alternate Solution}
Alternatively, we can also determine $x$-evolution of $F^{NS}(x,t)$ if $\alpha$ and $\beta$ of eq. (\ref{eqn:ch4eq4}) satisfies additional relation $\beta=\alpha^{m}$ \cite{RAJSARMA1,RAJSARMA2}, where $m$ is a positive integer. In this case, for low $x$, we get from eq. (\ref{eqn:ch4eq4}),
\begin{equation}
\label{eqn:ch4eq28}
{F^{NS}(x,t)Y^{NS}(x)=\alpha[t X^{NS}(x)]^n+\alpha^{m}} 
\end{equation}
where eq. (\ref{eqn:ch4eq4A}) and eq. (\ref{eqn:ch4eq4B}) have been used.

Differentiating eq. (\ref{eqn:ch4eq28}) with respect to $\alpha$, we get
\begin{equation}
\label{eqn:ch4eq29}
{\alpha=-\left(\frac{1}{m}\right)^{\frac{1}{m-1}}t^{\frac{n}{m-1}}[X^{NS}(x)]^{\frac{mn}{m-1}}} 
\end{equation}
which forbids m=1.

Putting this value of $\alpha$ in eq. (\ref{eqn:ch4eq28}), we ultimately get
\begin{equation}
\label{eqn:ch4eq30}
{F^{NS}(x,t)Y^{NS}(x)=\left\{-\left(\frac{1}{m}\right)^{\frac{1}{m-1}}-\left(\frac{1}{m}\right)^{\frac{m}{m-1}}\right\}t^{\frac{mn}{m-1}}[X^{NS}(x)]^{\frac{mn}{m-1}}}  
\end{equation}
\begin{equation}
\label{eqn:ch4eq31}
{F^{NS}(x_0,t)Y^{NS}(x_0)=\left\{-\left(\frac{1}{m}\right)^{\frac{1}{m-1}}-\left(\frac{1}{m}\right)^{\frac{m}{m-1}}\right\}t^{\frac{mn}{m-1}}[X^{NS}(x_0)]^{\frac{mn}{m-1}}}  
\end{equation}
we get,
\begin{equation}
\label{eqn:ch4eq32}
{F^{NS}(x,t)=F^{NS}(x_0,t)\frac{Y^{NS}(x_0)}{Y^{NS}(x)}\frac{[X^{NS}(x)]^{\frac{mn}{m-1}}}{[X^{NS}(x_0)]^{\frac{mn}{m-1}}}}
\end{equation}
If $m$ is very large, $\frac{m}{m-1}\approx1$, then eq. (\ref{eqn:ch4eq32}) reduces to eq. (\ref{eqn:ch4eq11}) and equivalently eq. (\ref{eqn:ch4eq12}). However, unlike in the previous case, there is no additional condition of $n>0$.
Similarly in case of high $x$ too, we get the $x$-distribution from the alternative way, as in low $x$, similar to eq. (\ref{eqn:ch4eq19}) and eq. (\ref{eqn:ch4eq20}), but without the additional condition on $p$ i.e $p<0$. Thus unlike in the general case, the alternate solution cannot distinguish between low $x$ and high $x$ regimes, which is its limitation.

\section{Results:}
\label{secchap4:Results}
\subsection{Result on the basis of Particular solution:}
\label{subsecchap4:Particular solution result.}

 We obtained an interpolating function $H(x,t)$ from CCFR Neutrino data \cite{DKCPKD} viz., 
\begin{equation}
\label{eqn:ch4eq33}
{H(x,t)=0.382[-5.976x+0.996(1-x)]}
\end{equation}
such that for $x\rightarrow 0$ and $x\rightarrow 1$, it has desired behaviour of $n(x,t)$ and $p(x,t)$ respectively, yielding $n_{num}\le 0.3586$ and $p_{num}\ge -2.151$
Assuming these limits as process independent, we use them to study $x$-distribution in Figure \ref{fig:ch4fig1}(a-e) and Figure \ref{fig:ch4fig2}(a-e) for low and high $x$ respectively using the following equations that have been redefined from eq. (\ref{eqn:ch4eq11}) and eq. (\ref{eqn:ch4eq19}) respectively as  
\begin{equation}
\label{eqn:ch4eq34}
{{F^{NS}(x,t)=F^{NS}(x_0,t)\frac{Y^{NS}(x_0)}{Y^{NS}(x)}\frac{[X^{NS}(x)]^{H(x,t)}}{[X^{NS}(x_0)]^{H(x_0,t)}}}}
\end{equation}
\begin{equation}
\label{eqn:ch4eq34A}
{{F^{NS}(x,t)=F^{NS}(x_0,t)\frac{Y^{{\prime}^{NS}}(x_0)}{Y^{{\prime}^{NS}}(x)}\frac{[X^{{\prime}^{NS}}(x)]^{H(x,t)}}{[X^{{\prime}^{NS}}(x_0)]^{H(x_0,t)}}}}
\end{equation}

We compare our results with CCFR data \cite{WCLEUNG,WJSELIGMANETAL} as well as GRV(98) \cite{GRV,GRV98} and MRST LO 2001 \cite{MRST11,MRST22,MRSTDURHAM} exact results. For getting the exact results, we take the non-singlet structure function in leading order as \cite{GKREYA},
\begin{equation}
\label{eqn:ch4eq35}
{xF_3(x,t)=-x(\overline{u}+\overline{d})+x(d+u)[|v_{ud}|^2+|v_{cd}|^2]+2xs[|v_{us}|^2+|v_{cs}|^2]-2xc}.
\end{equation}
where $u=u(x,t)$ etc. and $v_{ud}$, $v_{us}$ etc. are the relevant CKM matrix elements \cite{PDG}. We have taken $v_{ud}\approx 0.97$, $v_{us}\approx 0.22$, $v_{cd}\approx 0.22$ and $v_{cs}\approx 0.97$.

For low $x$ (Figure \ref{fig:ch4fig1}(a-e)), our preditions are found to be tallying very well (Within $3 \%$ limit) with both data and exact results in the lower part of the $x$ spectrum, but it does not match so well in the upper part of the $x$ spectrum. 

For high $x$ (Figure \ref{fig:ch4fig2}(a-e)), on the other hand, our predictions are closer to both data and exact results (Within $5 \%$ limit) in the extreme lower part of the $x$ spectrum. In the middle part of the $x$ spectrum, coincidence between our predictions and both data as well as exact results is not so well as expected. For extreme  higher parts of the $x$ spectrum, there is excellent coincidence between our predictions and both data and exact results (Specially in the region $0.60<x<0.75$).

The deviations of our predictions from data and exact results is presumably due to the neglect of finite $x$ corrections coming from the higher derivatives of $\frac{\partial F^{NS}(x,t)}{\partial x}$ in the Taylor approximation (Neglected in eq. (3) and eq.(35) of Ref.\cite{DKCPKD}). It may also be partly due to the neglect of explicit $x$ and $t$ dependence of $n(x,t)$ and $p(x,t)$ in developing the seperate formalism of low and high $x$.

\subsection{Result on the basis of General solution:}
\label{subsecchap4: General solution result.}

We redefine eq. (\ref{eqn:ch4eq24}) and eq. (\ref{eqn:ch4eq25})  as

\begin{equation}
\label{eqn:ch4eq36}
{F^{NS}(x,t)=F^{NS}(x_0,t)\frac{Y^{NS}(x_0)}{Y^{NS}(x)}\frac{t^{H(x,t)}}{t^{H(x_0,t)}}\frac{[X^{NS}(x)]^{H(x,t)}}{[X^{NS}(x_0)]^{H(x_0,t)}}}.    
\end{equation}

\begin{equation}
\label{eqn:ch4eq37}
{F^{NS}(x,t)=F^{NS}(x_0,t)\frac{Y^{{\prime}^{NS}}(x_0)}{Y^{{\prime}^{NS}}(x)}\frac{t^{H(x,t)}}{t^{H(x_0,t)}}\frac{[X^{{\prime}^{NS}}(x)]^{H(x,t)}}{[X^{{\prime}^{NS}}(x_0)]^{H(x_0,t)}}}.    
\end{equation}

which are valid for low $x$ and high $x$ regions respectively.

\begin{figure}
\begin{center}
\includegraphics[height=12cm,width=15cm]{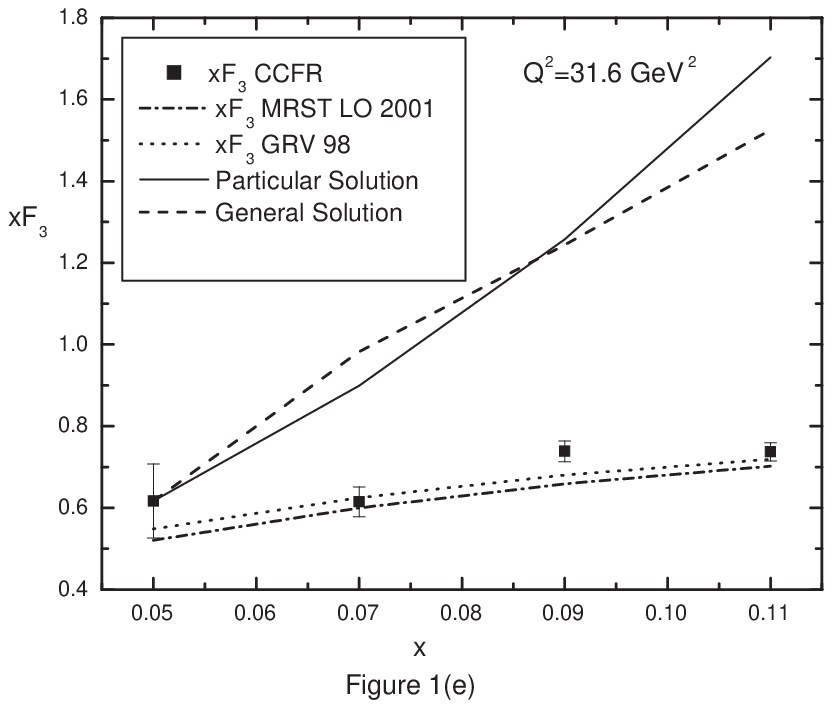}
\end{center}
\caption[ $x F_3$ versus $x$ in bins of $Q^2$ obtained from eq. (\ref{eqn:ch4eq34}) and eq. (\ref{eqn:ch4eq36}) for low $x$ values. This is compared with the GRV98  and MRST2001 exact results calculated using eq. (\ref{eqn:ch4eq35}).This is also compared with CCFR data.]{(a-e) $x F_3$ versus $x$ in bins of $Q^2$ obtained from eq. (\ref{eqn:ch4eq34}) (Solid line) and eq. (\ref{eqn:ch4eq36}) (Dashed curve) for low $x$ values. This is compared with the GRV98 \protect\cite{GRV,GRV98} (Dotted curve) and MRST2001 \protect\cite{MRST11,MRST22,MRSTDURHAM} (Dashed dotted curve) exact results calculated using eq. (\ref{eqn:ch4eq35}). This is also compared with CCFR data \protect\cite{WCLEUNG,WJSELIGMANETAL}.}
\label{fig:ch4fig1}
\end{figure}

\begin{figure}
\begin{center}
\includegraphics[height=12cm,width=15cm]{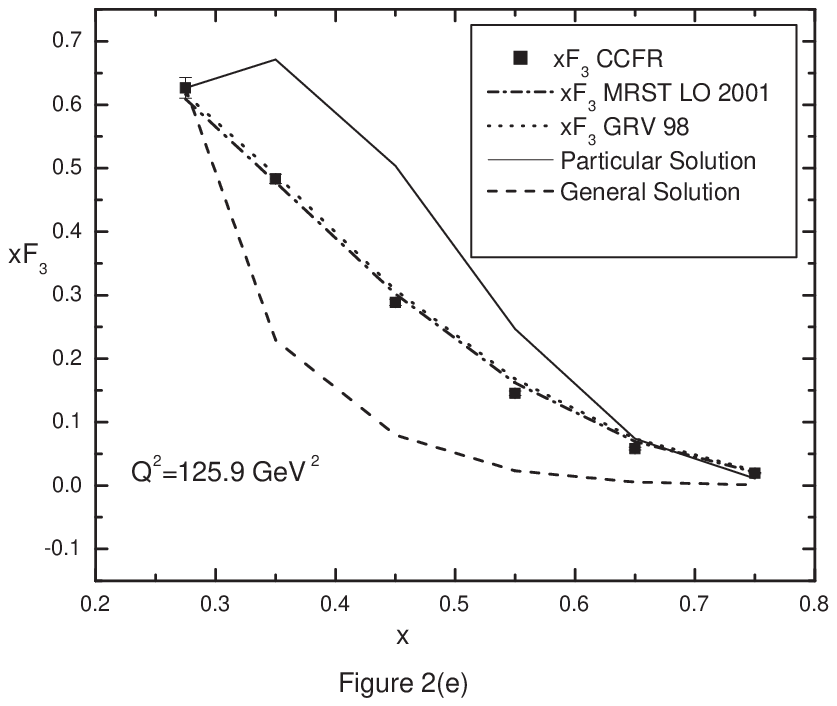}
\end{center}
\caption[ $x F_3$ versus $x$ in bins of $Q^2$ obtained from eq. (\ref{eqn:ch4eq34A}) and eq. (\ref{eqn:ch4eq37}) for high $x$ values. This is compared with the GRV98  and MRST2001 exact results calculated using eq. (\ref{eqn:ch4eq35}).This is also compared with CCFR data.]{(a-e) $x F_3$ versus $x$ in bins of $Q^2$ obtained from eq. (\ref{eqn:ch4eq34A}) (Solid line) and eq. (\ref{eqn:ch4eq37}) (Dashed curve) for high $x$ values. This is compared with the GRV98 \protect\cite{GRV,GRV98} (Dotted curve) and MRST2001 \protect\cite{MRST11,MRST22,MRSTDURHAM} (Dashed dotted curve) exact results calculated using eq. (\ref{eqn:ch4eq35}). This is also compared with CCFR data \protect\cite{WCLEUNG,WJSELIGMANETAL}.}
\label{fig:ch4fig2}
\end{figure}

On the basis of the above equations, we study the $x$-distribution in Figure \ref{fig:ch4fig1}(a-e) and Figure \ref{fig:ch4fig2}(a-e) for low $x$ and high $x$ respectively. The dash line in the above set of figures represents the result obtained on the basis of  general solution.

\begin{figure}
\begin{center}
\includegraphics[height=10cm,width=12cm]{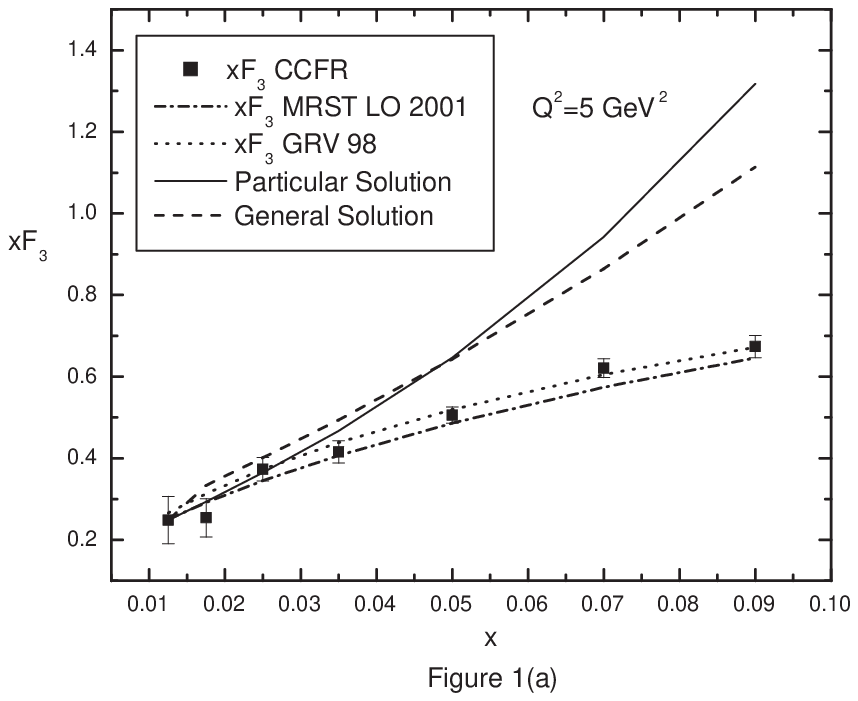}
\includegraphics[height=10cm,width=12cm]{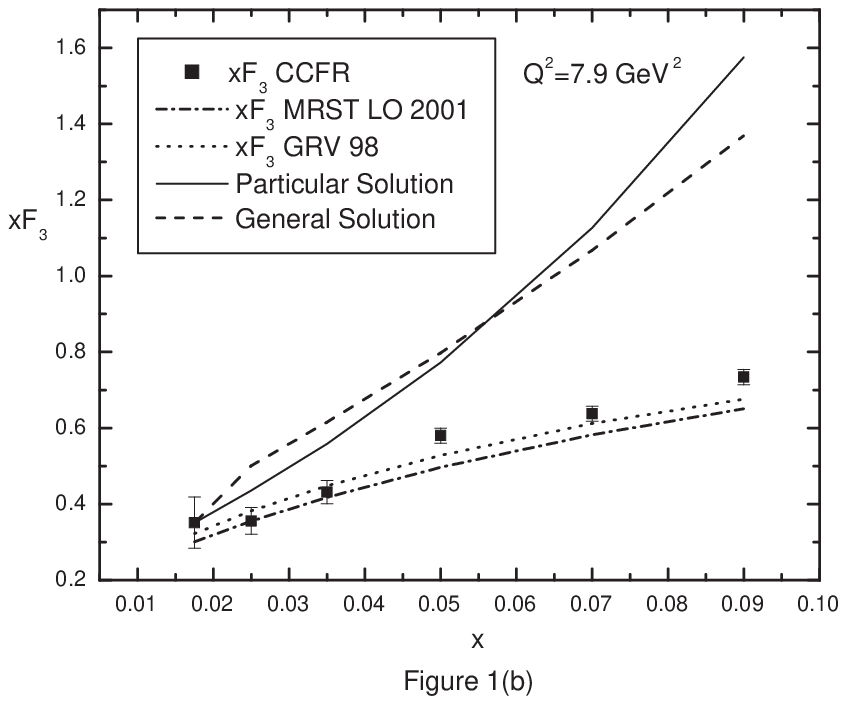}\\
\end{center}
\end{figure}

\begin{figure}
\begin{center}
\includegraphics[height=10cm,width=12cm]{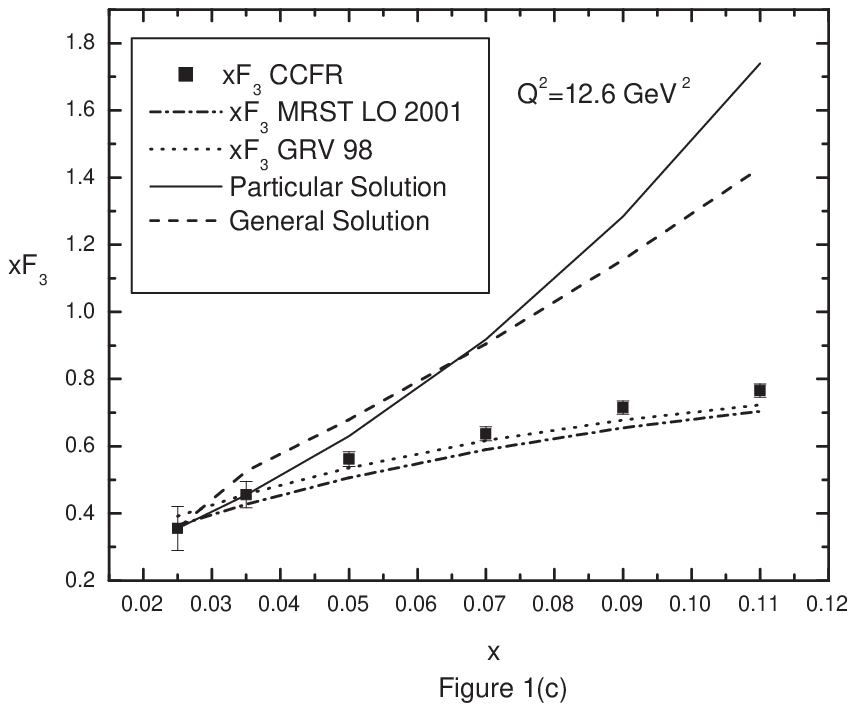}
\includegraphics[height=10cm,width=12cm]{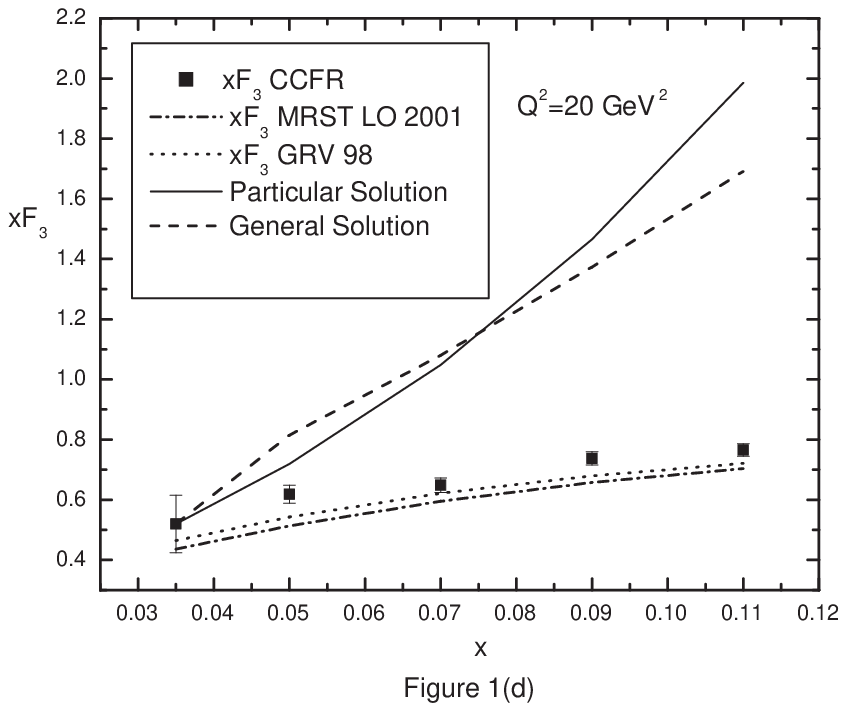}\\
\end{center}
\end{figure}

\begin{figure}
\begin{center}
\includegraphics[height=10cm,width=12cm]{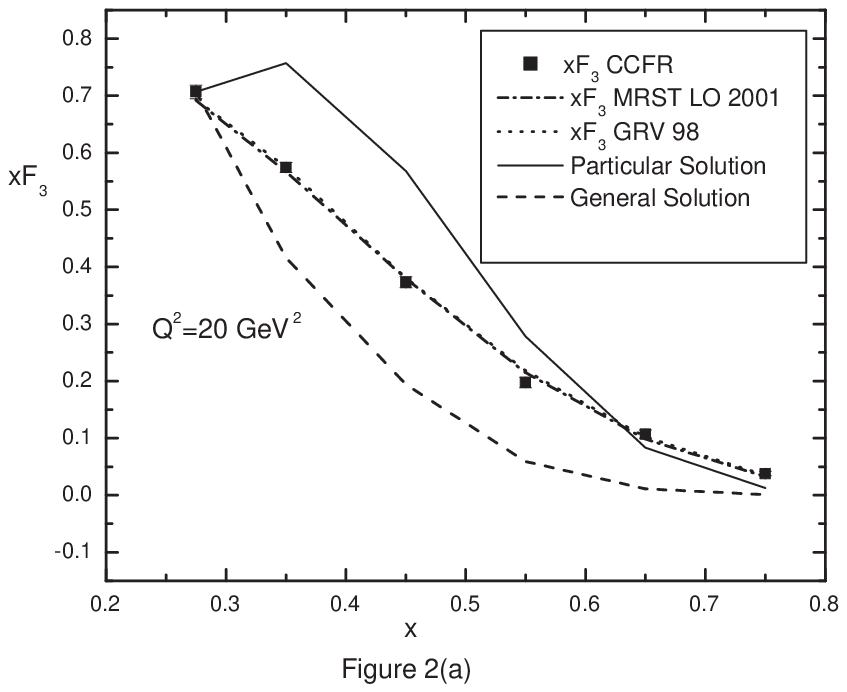}
\includegraphics[height=10cm,width=12cm]{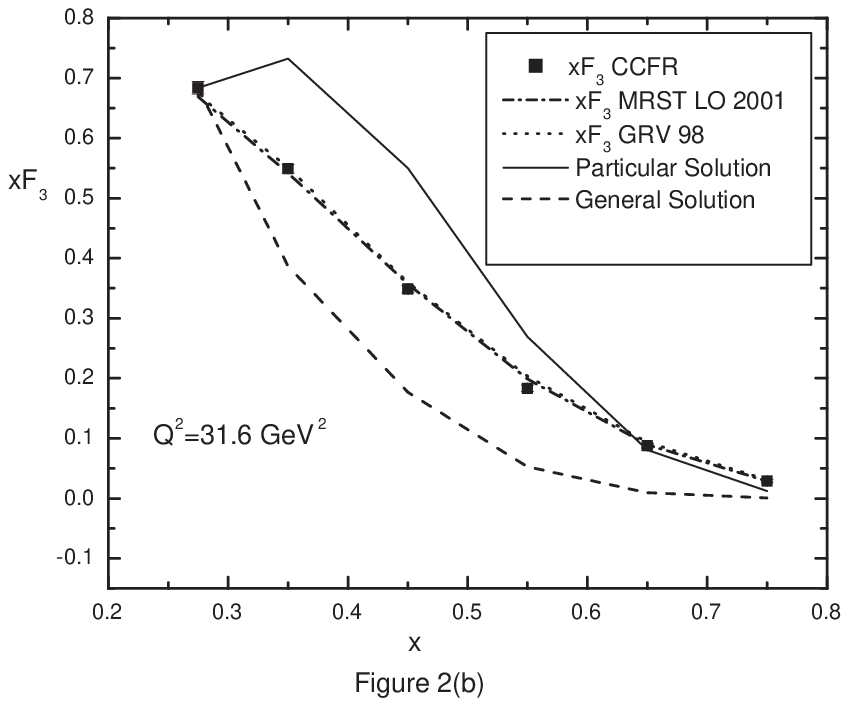}\\
\end{center}
\end{figure} 

\begin{figure}
\begin{center}
\includegraphics[height=10cm,width=12cm]{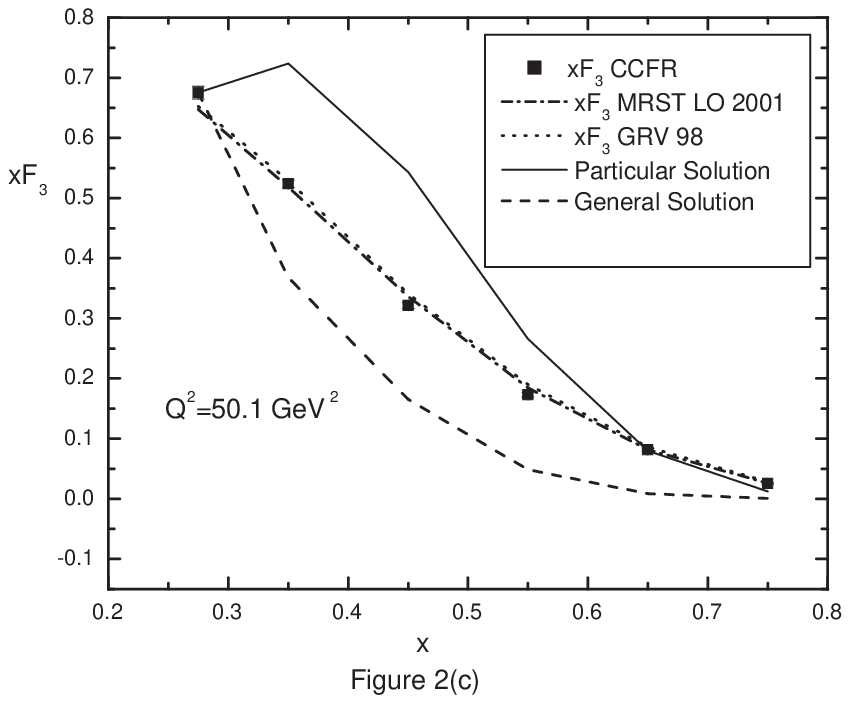}
\includegraphics[height=10cm,width=12cm]{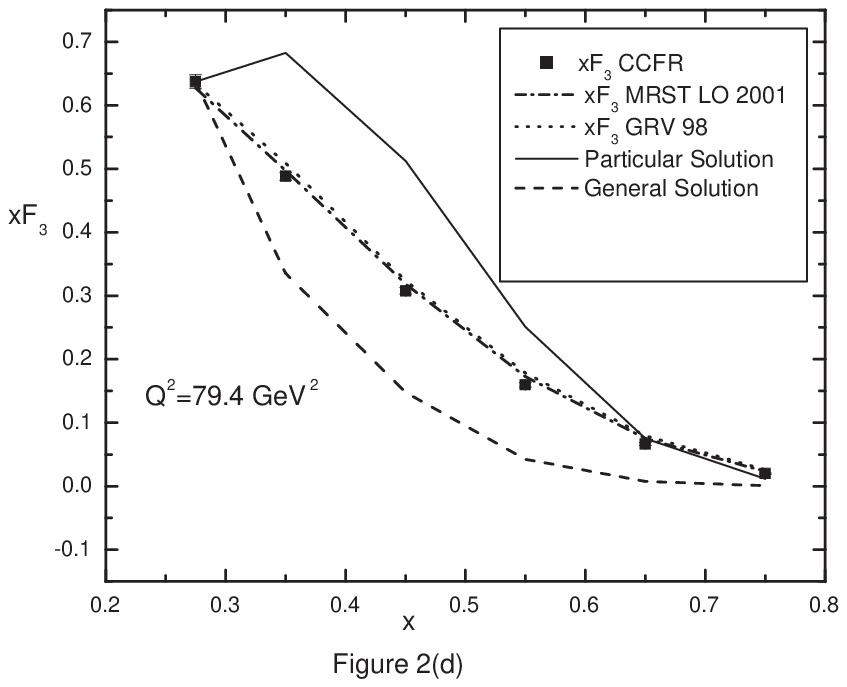}\\
\end{center}
\end{figure}

For low $x$, our predictions based on the particular and the general solution does not differ appreciably in the lower part of $x$-spectrum, but the general solution is found to match better with both data and exact results in the upper part of $x$ spectrum. On the other hand, for high $x$, the general solution matches better than particular solution with both data and exact results in the lower part of $x$ spectrum, but it is found that the opposite is true for higher part of $x$-spectrum.

Thus we find that acceptance of explicit $x$ and $t$-dependence of $n(x,t)$ and $p(x,t)$ gives better result in the higher $x$ spectrum
of low $x$ and lower $x$ spectrum of high $x$.

\subsection{Incorporation of large $x$ effect:}
\label{subsecchap4:large x effect.}
\begin{figure}
\begin{center}
\includegraphics[height=12cm,width=15cm]{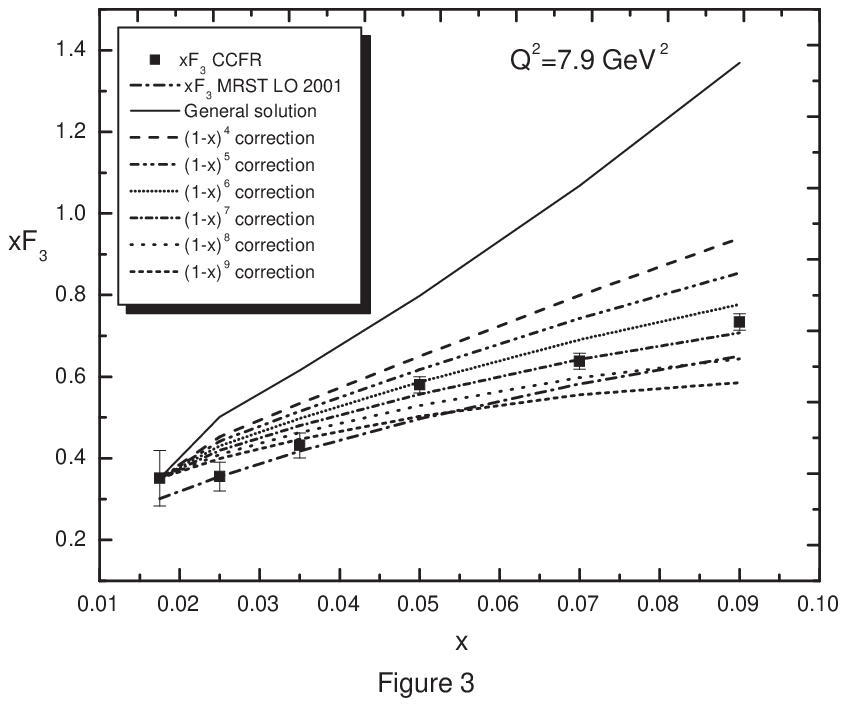}
\end{center}
\caption[ Modified $x F_3$ versus $x$ for fixed $Q^2=7.9$  $GeV^2$ simulated by an empirical multiplicative damping factor  $(1-x)^r$ in the structure function (General Solution) in eq. (\ref{eqn:ch4eq36}) for low $x$ values. This is compared with the  MRST2001 exact results calculated using eq. (\ref{eqn:ch4eq35}) as well as CCFR data.]{ Modified $x F_3$ versus $x$ for fixed $Q^2=7.9$  $GeV^2$ simulated by an empirical multiplicative damping factor  $(1-x)^r$ in the structure function (General Solution) in eq. (\ref{eqn:ch4eq36}) for low $x$ values. This is compared with the MRST2001 \protect\cite{MRST11,MRST22,MRSTDURHAM} (Dashed dotted curve) exact results calculated using eq. (\ref{eqn:ch4eq35}) as well as CCFR data \protect\cite{WCLEUNG,WJSELIGMANETAL} .}
\label{fig:ch4fig3}
\end{figure}

\begin{figure}
\begin{center}
\includegraphics[height=12cm,width=15cm]{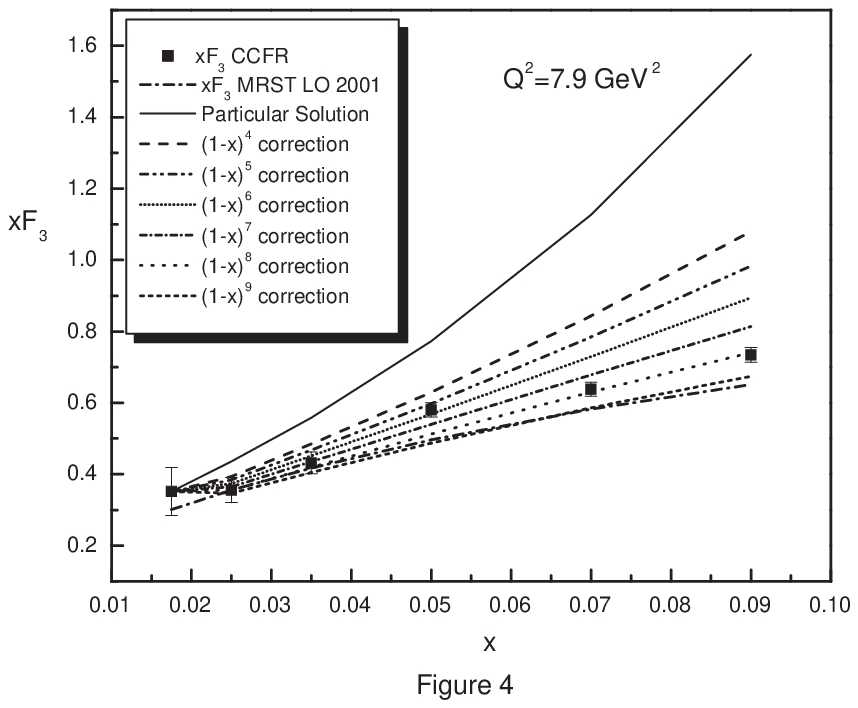}
\end{center}
\caption[ Modified $x F_3$ versus $x$ for fixed $Q^2=7.9$  $GeV^2$ simulated by an empirical multiplicative damping factor  $(1-x)^r$ in the structure function (particular Solution)  in eq. (\ref{eqn:ch4eq34}) for low $x$ values. This is compared with the  MRST2001 exact results calculated using eq. (\ref{eqn:ch4eq35}) as well as CCFR data.]{ Modified $x F_3$ versus $x$ for fixed $Q^2=7.9$  $GeV^2$ simulated by an empirical multiplicative damping factor  $(1-x)^r$ in the structure function (particular Solution)  in eq. (\ref{eqn:ch4eq34}) for low $x$ values. This is compared with the MRST2001 \protect\cite{MRST11,MRST22,MRSTDURHAM} (Dashed dotted curve) exact results calculated using eq. (\ref{eqn:ch4eq35}) as well as CCFR data \protect\cite{WCLEUNG,WJSELIGMANETAL} .}
\label{fig:ch4fig4}
\end{figure}
For low $x$ , we find in Subsection  \ref{subsecchap4: General solution result.} that predictions based on the most general solution have shoot up the CCFR neutrino data as well as exact results in the higher part of $x$ -spectrum. This overshooting can be substantially reduced by the introduction of large $x$ effect. Such finite $x$ corrections, which may be considered as the conglomeration of the terms neglected while getting  eq. (\ref{eqn:ch4eq2}), can effectively be simulated by an empirical multiplicative damping factor  $(1-x)^r$, ($r$ is positive) in the structure function (General Solution) in eq. (\ref{eqn:ch4eq36}). We have varied $r$ from 4 to 9 and found $r$ around 7 to yield the best fit to the data and exact result. This is clearly shown in Figure \ref{fig:ch4fig3}, where we have plotted this modified structure function at different $x$ for fixed $Q^2=7.9$  $GeV^2$. Similarly, we multiply the structure function (particular Solution) in eq. (\ref{eqn:ch4eq34}) by multiplicative damping factor  $(1-x)^r$ and plot this modified structure function versus $x$ for fixed $Q^2=7.9$  $GeV^2$ in Figure \ref{fig:ch4fig4}. This time we find the value of $r$ around 8 gives the best fit to the data, clearly pinpointing the supremacy of the general solution over the particular solution. It is pertinent to note that the modified structure function damps considerably in the higher part of $x$ -spectrum (of low $x$) only due to the damping factor. In the lower part of the $x$ spectrum (of low $x$), there is very little damping because $(1-x)^r \rightarrow 1$ when $x \rightarrow 0$.

\subsection{Comparison with other works:}
\label{subsecchap4:other works.}

Let us now discuss how the present work differs from that of Ref.\cite{JKSDKCGKM}, \cite{RAJSARMA1} and \cite{RAJSARMA2}. In Ref.\cite{JKSDKCGKM}, Taylor-approximated leading order coupled DGLAP equation for singlet ($F_2^S(x,t)$) and gluon distribution ($G(x,t)$) were solved using the relationship $G(x,t)=k(x)F_2^S(x,t))$ and the particular solution $v=\alpha u+\beta$ ($\alpha$ and $\beta$ are constants)
implying $n(x,t)=1$ in the general solution of the present formalism (eq. (\ref{eqn:ch4eq4})). The results are then compared with deuteron data using the relation $F_2^d(x,t)=\frac{5}{9}F_2^S(x,t)$.

In Ref.\cite{RAJSARMA1}, a similar analysis was done for deuteron structure function using the additional relation $\beta=\alpha^2$. In Ref.\cite{RAJSARMA2}, the analysis was further extended to Next-to-Leading order for deuteron, proton and neutron structure function  assuming the relationship $\beta=\alpha^m (m=2,3,4,5.....)$ implying that $x$ evolution of the formalism is not unique. In the present work, we have made the formalism much more general abandoning the linearity of $u$ but still preserving that of $v$ (which is equivalent to structure function) in the general solution.

\section{Comments and conclusions:}
\label{secchap4:Comments and conclusions}
To conclude, we have shown that the $x$-distribution of the non-singlet structure function obtained from Leading Order Taylor-approximated DGLAP equations are compatible with data \cite{WCLEUNG,WJSELIGMANETAL} as well as exact results for low and high $x$ \cite{MRST11,MRST22,MRSTDURHAM,GRV,GRV98} in selected parts of the $x$ spectrum. We have also shown how the introduction of damping factor considerably brings the modified structure function closer to exact resuts in case of low $x$. We have also discussed how the present formalism is an improvement and generalization over those of Ref.\cite{JKSDKCGKM,RAJSARMAEUR,RAJSARMA1,RAJSARMA2}.

\end{document}